# Participatory Approaches in AI Development and Governance |
# A Principled Approach

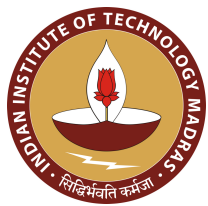
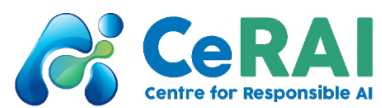

**About Vidhi Centre for Legal Policy**

**About the Centre for Responsible AI**

Vidhi Centre for Legal Policy ('Vidhi') is an independent think-tank undertaking legal research to make better laws and improve governance for the public good. We do this through high quality, peer reviewed original legal research, engaging with the Government of India, State governments and other public institutions to both inform policymaking and to effectively convert policy into law. Since 2013, Vidhi has worked with over 20 different ministries in the Government of India, 9 State Governments, the Supreme Court of India, 14 regulators and public institutions for different projects, including 77 enactments of binding law or policy. During this period, Vidhi has also produced over 369 pieces of original legal research.

Website: www.vidhilegalpolicy.in

Centre for Responsible AI (CeRAI) is a virtual interdisciplinary research centre at IIT Madras focused on fundamental and applied research in Responsible AI. Our research includes technical research on making AI models and products fair, understandable and safe, policy analysis on governance and regulation of AI, and substantial training and outreach programs addressing a wide audience ranging from primary school children to senior executives.

Website: https://cerai.iitm.ac.in/

# About the Authors

## Vidhi

**Ameen Jauhar** was a Senior Resident Fellow and Team Lead at the Centre for Applied Law and Technology (ALTR) at Vidhi.

**Aditya Phalnikar** is a Research Fellow at the Research Director's Office at Vidhi.

**Dhruv Somayajula** was a Senior Resident Fellow at the Centre for Applied Law and Technology (ALTR) at Vidhi.

## CeRAI

**Ambreesh Parthasarathy** is a Post Baccalaureate Fellow at the Centre for Responsible AI (CeRAI) at IITM.

**Prof. Balaraman Ravindran** is the head of the Centre for Responsible AI (CeRAI) at IITM.

**Gokul Krishnan** is a Research Scientist at the Centre for Responsible AI (CeRAI) at IITM.

The authors would also like to thank Sanjay Karanth, Ankit Bose, Raj Shekhar, Ikran Ali Abdirahman, Jai Vipra, Urvashi Aneja, Tarunima Prabhakar, Jibu Elias, Vidushi Marda, Harish Guruprasad, Gokul Krishnan, Susanna Pirttikangas and Lauri Lovén for their participation in a consultation on an earlier draft of this paper.

Vidhi would also like to thank the Lal Family Foundation for their generous funding to support Vidhi's research.

Errors in the paper, if any, are the sole responsibility of the authors.

# Table of Contents

# Executive Summary

The widespread adoption of Artificial Intelligence (AI) technologies in the public and private sectors has resulted in them significantly impacting the lives of people in new and unexpected ways. In this context, it becomes important to inquire how their design, development and deployment takes place. Upon this inquiry, it is seen that persons who will be impacted by the deployment of these systems have little to no say in how they are developed. Seeing this as a lacuna, this research study advances the premise that a participatory approach is beneficial (both practically and normatively) to building and using more responsible, safe, and human-centric AI systems. Normatively, it enhances the fairness of the process and empowers citizens in voicing concerns to systems that may heavily impact their lives. Practically, it provides developers with new avenues of information which will be beneficial to them in improving the quality of the AI algorithm.

The paper advances this argument *first*, by describing the life cycle of an AI system; *second*, by identifying criteria which may be used to identify relevant stakeholders for a participatory exercise; and *third*, by mapping relevant stakeholders to different stages of AI lifecycle.

This paper forms the first part of a two-part series on participatory governance in AI. The second paper will expand upon and concretise the principles developed in this paper and apply the same to actual use cases of AI systems.

# Introduction

In recent years, the world has witnessed astonishing progress in the field of AI and its uses. AI can be understood as a system that, given some data, can extract and learn the intrinsic rules and features present in it and use this learning to perform tasks. This is vastly different from conventional automation (using ICT or Information and Communication Technologies) where the rules for decision making are pre-determined. This aspect of extraction and learning is what makes AI unique. The rollout of AI systems generally is preceded by the following questions: (a) Is AI necessary to solve the problem? (b) Is there literature or work supporting the solution logic? (c) Is there high-quality data available? (d) Is the solution feasible? An AI system is developed, in stages, after these questions have been dealt with. The AI system lifecycle phases are the design phase, the development phase, and the deployment phase.[1] The foregoing stages are neither exhaustive nor sharply defined, often blending into each other based on the circumstances in which the AI systems are being developed and deployed. In addition to these stages leading up to deployment, an AI system's life cycle includes continuous monitoring and feedback regarding its performance at its functions.[2]

Several fields, ranging from the essential to the mundane, are already relying on AI systems for both core and penumbral functions. The ability of AI systems to be customised to carry out specific roles within a field, along with its inherent computational capabilities, promises fantastic efficiencies. Certain AI systems like chatbots and voice assistants can digitise existing human functions. We also see AI systems designed to complement and assist existing human functions for better and more efficient functionality. Some forms of AI systems operate independently of human oversight, be it through recommender algorithms, generative models, or as forecasting and modelling systems that seek to identify newer insights by analysing troves of existing data. The staggering potential of AI systems is complemented by their diverse roles and use-cases. However, with its ever-expanding usage, there are parallel concerns around its risks of potential or actual harm. For instance, biased outputs of an AI system can cause real world harm to individuals who are at the receiving end. An example of this is where facial recognition technology (**FRT**) has misidentified individuals resulting in their arrests, prosecution, and even incarceration. The conversation around mitigating AI harms and maximising its benefits is covered within the field of AI ethics promoting the responsible, safe and trustworthy use of AI systems.

A key discussion that has emerged in the larger discourse of responsible AI, is the need for human-centricity, and how to ensure AI design and deployment favours and not impedes core human rights and values. Herein, Participatory AI (**PAI**) is an idea that has gained increasing traction. Predicated on rejigging the dynamics between AI developers and deployers on the one hand, and the affected persons and users on the other, it fundamentally seeks to create a balance among competing interests but affording decision-making authority to more and diverse groups and individuals.

---

[1] Daswin De Silva and Damminda Alahakoon, 'An Artificial Intelligence Life Cycle: From Conception to Production' (2022) 6(3) Patterns 1, 4.
[2] ibid.

This paper is one of a two-part study. It lays out the general idea of participation, and how it has been utilised in other sectors. It then provides a general overview of how AI has been used in both the public and private sectors, and the consequent problems that have arisen due to a lack of participatory approach. The third part deals with operationalising PAI to address these problems. Specifically, it will provide theoretical context and ideas for:

(a) breaking down the different stages of the AI life cycle (i.e., design, development, and deployment);

(b) how stakeholders should be identified;

(c) how stakeholders are distributed across these different stages to facilitate meaningful engagement and inputs from them; and

(d) the rules of collation, i.e., how inputs from participants should be collated and synthesised, and their incorporation into the actual lifecycle of an AI system.

The last section will offer concluding remarks.

# Participation Elsewhere

The core tenets of PAI derive from larger discussions around participatory governance. As briefly mentioned in the introduction, the intention behind such frameworks is to ensure greater transparency and authority to stakeholders who have traditionally been on the receiving end of governance frameworks.

This section gives a snapshot of how participatory governance has been used in India and abroad, for legislation and policy design and development.

| Area of governance | Degree of Participation |
|---|---|
| Nagoya Protocol | The Nagoya Protocol on Access to Genetic Resources and the Fair and Equitable Sharing of Benefits Arising from their Utilisation requires member states to ensure that benefits arising from the use of traditional knowledge must be fairly shared, on mutually agreed terms, with the peoples, local communities, and indigenous peoples holding such traditional knowledge.[3] This includes obtaining prior informed consent of all stakeholders, i.e., local communities and indigenous peoples' active involvement and approval to access the traditional knowledge associated with genetic resources.[4] |
| Land Acquisition Act | The Right to Fair Compensation and Transparency in Land Acquisition, Rehabilitation and Resettlement Act, 2013 ("**Land Acquisition Act**") was enacted to offer fair compensation to owners of land acquired for public purposes under this Act.[5] The Act requires the preparation of a social impact assessment study in collaboration with all affected families, to be discussed publicly.[6] A key pillar of the Act is the role of prior consent for various kinds of acquisitions, including the consent of the Gram Sabha for any land acquisition in Scheduled Areas, or the consent of 80% of affected families for acquisitions for private companies for public purposes.[7] |
| Forest Rights Act | Section 6 of the Scheduled Tribes and Other Traditional Forest Dwellers (Recognition of Forest Rights) Act, 2006 provides for stakeholder collaboration and input as far as the procedure for |

---

[3] Nagoya Protocol, Art 5(5).
[4] Nagoya Protocol, Art 7.
[5] The Right to Fair Compensation and Transparency in Land Acquisition, Rehabilitation and Resettlement Act, 2013 (Land Acquisition Act).
[6] Land Acquisition Act, s 5.
[7] Land Acquisition Act, ss 2(2) and 41.

|  |  |
|---|---|
|  | vesting forest rights is concerned.[8] The Gram Sabha has been preferred as the statutory institution due to its evident participatory and democratic nature. Further, there are provisions for two levels of appeals that can be preferred by the affected communities/party if they feel that the institutional decisions against their claims are unjust. |
| Labour Laws | Indian labour law[9] places importance on collaboration between the employer and the employee so that neither is unfairly exploited nor strong-armed by the other. Legislations such as the Industrial Disputes Act, 1947[10] (which establishes a framework for the registration and formation, apart from the formalities for dispute resolution) and the Trade Unions Act, 1926[11] (an important law giving legal credence to trade unions, and outlining their rights and responsibilities) are both examples of special legislations which aim to encourage participation and amicable settlement of disputes. Other laws and regulations that impact collective bargaining in India include the Minimum Wages Act, 1948,[12] the Payment of Bonus Act, 1965,[13] and the Factories Act, 1948.[14]<br><br>Under the Industrial Disputes Act, 1947 collective bargaining practices have been recognised and regulated. As far as the Trade Unions Act, 1926 is concerned, the provision for appeal against refusal for registration of a trade union (Section 11),[15] and the mandatory requirement for the previous publication of any rules made under this Act by the appropriate Government (Section 30)[16] are indicative of the participatory nature of the Act. |
| Town Planning | Development and design of land use in urban areas is an important task that falls before all governments. Increasingly, there has been a push towards direct or indirect public participation in this process of town planning - which ultimately leads to efficient and sustainable cities. Participatory governance is a key aspect of town planning in Japan (*machizukuri*), emphasising the active involvement of community members in decision-making processes related to urban |

---

[8] The Scheduled Tribes and Other Traditional Forest Dwellers (Recognition of Forest Rights) Act, 2006, s 6, No. 2, Acts of Parliament, 2007 (India).
[9] Although the President has assented to the new labour laws (comprising of the Code on Wages 2019, Occupational Safety, Health and Working Conditions Code 2020, Code on Social Security 2020, and the Industrial Relations Code 2020), none of these laws have been completely notified. Since the old laws continue to be in force, they have been referred to for the purposes of this discussion.
[10] The Industrial Disputes Act 1947, No. 14, Acts of Parliament, 1947 (India).
[11] The Trade Unions Act 1926, No. 16, Acts of Parliament, 1926 (India).
[12] The Minimum Wages Act 1948, No. 11, Acts of Parliament, 1948 (India).
[13] The Payment of Bonus Act 1965, No. 21, Acts of Parliament, 1965 (India).
[14] The Factories Act, 1948 No. 63, Acts of Parliament, 1948 (India).
[15] The Trade Unions Act, 1926, s 11, No. 16, Acts of Parliament, 1926 (India).
[16] The Trade Unions Act, 1926, s 30, No. 16, Acts of Parliament, 1926 (India).

| | development. It recognises that residents are the experts of their own communities, possessing valuable knowledge, experiences, and insights that can contribute to more effective and sustainable planning outcomes. Their participation can manifest as community meetings, workshops, focus groups, or online platforms where residents can voice their concerns, propose ideas, and collaborate with local authorities and other stakeholders. |
|---|---|

**Table 1:** *The table summarises how participatory approaches have been used in other sectors.*

# Theoretical Overview of PAI

The design, development and deployment of AI systems has been embraced by the state in its public functions and by private actors. For example, AI systems such as FRT are being used for law enforcement in Tamil Nadu (FaceTagr), Punjab (PAIS), Uttar Pradesh (Trinetra) and New Delhi (AI Vision).[17] Additionally, FRT is being used for delivery of services and increasing efficiency in governance such as the Real Time Digital Authentication of Identity project to authenticate pensioners in Telangana,[18] the DigiYatra authentication project being carried out in select airports across India,[19] and the Face Matching Technology system adopted by the Central board for Secondary Education to provide access to academic documents by authenticating a student's identity through FRT.[20] Elsewhere, the Indian government has embraced the use of AI-enabled chatbots and voice assistants as seen in the 'Unified Mobile Application for New-age Governance' or 'UMANG' app,[21] the Corona Chatbot released by MyGov for citizen awareness during the pandemic,[22] and the 'Aaple Sarkar' chatbot released to inform citizens of the services provided by the Maharashtra government.[23] AI systems have also been deployed for analytics, such as the Investigation Tracking System for Sexual Offences (**ITSSO**). The ITSSO leverages data from the Crime & Criminal Tracking Network and System for filing First Information Reports (**FIRs**) and final reports, monitoring resolution of sexual offences cases in real time and generating crime heat maps or hot spots for appropriate deployment of resources.[24] AI systems are also being used for fraud detection by the Indian government. The Goods and Services Tax Network has commenced using an AI system named the 'Business

---

[17] Divya Chandrababu, 'Facial recognition system of Tamil Nadu police stirs privacy row' (Hindustan Times 10 December 2022) <https://www.hindustantimes.com/india-news/facial-recognition-system-of-tn-police-stirs-privacy-row-101670614143523.html> accessed 28 September 2023; Gopal Sathe, 'Cops In India Are Using Artificial Intelligence That Can Identify You In a Crowd' (Huffpost 16 August 2018) <https://www.huffpost.com/archive/in/entry/facial-recognition-ai-is-shaking-up-criminals-in-punjab-but-should-you-worry-too_in_5c107639e4b0a9576b52833b> accessed 28 September 2023; 'Staqu launches TRINETRA, an AI app for UP Police Department' (Deccan Chronicle December 29 2018) <https://www.deccanchronicle.com/technology/in-other-news/291218/staqu-launches-trinetra-an-ai-app-for-up-police-department.html> accessed 28 September 2023; Varsha Bansal, 'The Low Threshold for Face Recognition in New Delhi' (Wired 21 August 2022) <https://www.wired.co.uk/article/delhi-police-facial-recognition> accessed 28 September 2023.
[18] 'Telangana government leveraging the power of AI and ML for pensioners' (IndiaAI 27 October 2022) <https://indiaai.gov.in/case-study/telangana-government-leveraging-the-power-of-ai-and-ml-for-pensioners> accessed 28 Septemeber 2023.
[19] Saurabh Sinha, 'DigiYatra Roll out: Your face will now be an ID & domestic boarding card at Delhi, Bengaluru and Varanasi airports' (Times of India 5 December 2022) <https://timesofindia.indiatimes.com/business/india-business/digiyatra-rolled-out-your-face-will-now-be-your-id-and-domestic-ticket-at-delhi-bengaluru-and-varanasi/articleshow/95912778.cms> accessed 28 September 2023.
[20] 'CBSE introduces 'Facial Recognition System' for accessing digital academic documents of Class 10 and 12' (Hindustan Times 22 October 2020) <https://www.hindustantimes.com/education/cbse-introduces-facial-recognition-system-for-accessing-digital-academic-documents-of-class-10-and-12/story-XmoqbgNRCeVD9X91zzxFzM.html> accessed 28 September 2023.
[21] Nidhi Singal, 'International version of UMANG app launched by IT Minister Ravi Shankar Prasad' (Business Today 23 November 2020) <https://www.businesstoday.in/technology/news/story/international-version-of-umang-app-launched-by-it-minister-ravi-shankar-prasad-279392-2020-11-23> accessed 28 September 2023.
[22] 'How India's AI-enabled Corona Helpdesk is empowering citizens' (IndiaAI 11 August 2021) <https://indiaai.gov.in/article/how-india-s-ai-enabled-corona-helpdesk-is-empowering-citizens> accessed 28 September 2023.
[23] 'Government of Maharashtra launches Aaple Sarkar chatbot with Haptik' (Economic Times 5 March 2019) <https://economictimes.indiatimes.com/news/politics-and-nation/government-of-maharashtra-launches-aaple-sarkar-chatbot-with-haptik/articleshow/68268917.cms?from=mdr> accessed 28 September 2023.
[24] Ministry of Home Affairs, 'SheRaksha' (March 2019) <https://origin1504-mha.nic.in/sites/default/files/2022-09/WomenSafetyDivision_SheRakshaVol1_30052019pdf%5B1%5D.pdf>.

Intelligence and Fraud Analysis' to track tax evasion fraud using data mining to identify anomalous transactions and patterns.[25]

Similarly, private actors have incorporated AI systems in their functions for purposes such as fraud detection, predictive analytics, and training of employees.[26] ZestMoney, a prominent lending start-up, has incorporated AI technology based on machine-learning to assess creditworthiness of applicants using a wide range of consumer data and to conduct and risk assessments.[27] In the healthcare sector, SigTuple has launched its flagship AI system 'Manthana' to help pathologists and radiologists diagnose diseases using predictive and prescriptive algorithms. Similarly, Apollo Hospitals has launched the 'Clinical Intelligence Engine', a self-learning AI engine used by over 4000 doctors, relying on Machine Learning (**ML**) of a vast amount of health data to support clinical decisions and to assist with primary care and condition management.[28] Automakers such as MG Motor India and Mahindra & Mahindra have launched cars with AI-enabled features such as voice assistants and advanced driver-assistance systems.[29]

For how much it has been integrated into market products and governance systems, there is little to no discussion on how AI systems ought to incorporate a participatory approach to their entire lifecycle. This is especially critical given the (harmful) impact that AI systems can sometimes have on people's lives. As seen above, AI systems can make healthcare decisions, drive cars, and even help detect tax fraud. It is also concerning because they may not make these decisions accurately.[30]

Alternatively, we (as users) may not know the method by which an AI software has arrived at a particular decision. This is often called the 'black box problem'. This problem arises because AI systems learn from data using a variety of ways.[31] Vast datasets are utilised in order to train these systems after they have been designed. This includes the collection and refinement of data, developing test sets and training sets, and preserving these datasets for future uses.[32]

---

[25] Shakeel Maqbool, 'Think twice before tax fraud, the authorities are on to you with AI' (The Print 31 January 2023) <https://theprint.in/opinion/think-twice-before-tax-fraud-the-authorities-are-on-to-you-with-ai/1343045/> accessed 28 September 2023; Satyen Bordoloi, 'Government vs Tax fraud: The AI checkmate' (Sify 27 February 2023) <https://www.sify.com/ai-analytics/government-vs-tax-fraud-the-ai-checkmate/> accessed 28 September 2023.
[26] Poornima Nataraj, 'How are Indian banks adopting digitisation and AI?' (AnalyticsIndia, 11 April 2022) <https://analyticsindiamag.com/how-are-indian-banks-adopting-digitisation-and-ai/> accessed 28 September 2023.
[27] Srishti Deoras, 'A look at ZestMoney — infusing AI for creditworthiness and cardless EMI' (AnalyticsIndia 24 July 2017) <https://analyticsindiamag.com/look-zestmoney-infusing-ai-credit-worthiness-cardless-emi/> accessed 28 September 2023.
[28] 'Revolutionising healthcare. Apollo Hospitals launches AI-powered Clinical Intelligence Engine for doctors' (BusinessLine 7 February 2023) <https://www.thehindubusinessline.com/companies/apollo-hospitals-launches-ai-powered-clinical-intelligence-engine-for-doctors/article66478759.ece
> accessed 28 September 2023.
[29] Prasid Banerjee, 'AI set to steer cars in new direction' (Livemint 22 August 2021) <https://www.livemint.com/technology/tech-news/ai-set-to-steer-indian-cars-in-new-direction-11629652014913.html> accessed 28 September 2023.
[30] See Julia Angwin and others, 'Machine Bias' (ProPublica 23 May 2016) <https://www.propublica.org/article/machine-bias-risk-assessments-in-criminal-sentencing>. Their empirical analysis revealed that black people were much more likely to be incorrectly labelled as high risk, and white people were much less likely to do so. In other words, the discriminatory ability of the algorithm was low, which was most likely a result of biased data due to structural racism. COMPAS is a proprietary software, which also means that it is not possible to know the weightage that is given to different factors. See David Spiegelhalter, The Art of Statistics: Learning from Data (Pelican 2019). See also Kori Hale, 'A.I. Bias Caused 80% Of Black Mortgage Applicants To Be Denied' (Forbes 2 September 2021) <https://www.forbes.com/sites/korihale/2021/09/02/ai-bias-caused-80-of-black-mortgage-applicants-to-be-denied/?sh=3eaaa15736fe>.
[31] These include deep learning, reinforcement learning, and unsupervised learning or semi-supervised learning. Eduardo Morales and Hugo Jair Escalante, 'A brief introduction to supervised, unsupervised, and reinforcement learning' in Alejandro Torres-Garcia and others (eds.) Biosignal Processing and Classification using Computational Learning and Intelligence: Principles, Algorithms, and Applications (Academic Press 2021) 111.
[32] Digital Curation Centre, The Role of Data in AI Report for the Data Governance Working Group of the Global Partnership of AI (University of Edinburgh, 2020) accessed at <role-of-data-in-ai.pdf (gpai.ai)>.

However, the manner in which the AI system is trained in order to achieve the desired output values differs from model to model, and depends on the underlying algorithms, datasets used and the manner in which the model is trained on these datasets. A primary concern with such AI systems is their opacity. An increase in the complexity of AI algorithms provides for greater accuracy due to greater computational powers.[33] This allows for widespread use of AI systems, increasing automation in routine or instruction-specific tasks previously done by humans. Trials that show greater accuracy also spur confidence in the adoption of AI based applications, resulting in a drive to make AI systems more complex and more accurate in their results.

However, this increased complexity has a tangible impact on the interpretability or explainability of the functions of an algorithm, with humans facing challenges in understanding and interpreting the outputs provided by modern cutting-edge algorithms.[34] The loss of interpretability results in a greater inclination to trust the results delivered by the machine, also referred to as *automation bias*.[35] Further, there are scenarios where an algorithm may excel in its pre-launch evaluations, and yet may result in spurious results due to various factors when deployed for real-world use.[36] In the event of any bias, discrimination, exclusion or omissions due to actions of an uninterpretable AI system, this very complexity hurts attempts to understand the problematic components of the AI and results in a failure to identify areas within the AI's algorithmic process that need to be addressed.[37] The black box problem and automation bias, hence, result in a situation where, despite having great potential to impact lives (both individually and at a collective level), the public has no access to AI systems- neither in their design, nor in their development and deployment.

In this regard, PAI is a viable approach to increase meaningful participation of the community at large. It is centred around users, non-user affected persons and other stakeholders working with technical designers and developers in the design process.[38] The essential idea is that information that is available with the public (and which would be difficult for decision makers to access without involving the public) ought to be included in the decision-making process. This can be either through deliberative consensus building or preference aggregation (voting).[39]

---

[33] Tim Hwang, 'Computational Power and the Social Impact of Artificial Intelligence' (2019) accessed at <https://papers.ssrn.com/sol3/papers.cfm?abstract_id=3147971>

[34] Cynthia Rudin, 'Stop Explaining Black Box Machine Learning Models for High Stakes Decisions and Use Interpretable Models Instead' (2019) 1(5) Nature Machine Intelligence 1.

[35] Linda Skitka and others, 'Does automation bias decision-making?' (1999) 51 International Journal of Human-Computer Studies 991; Sarah Valentine, 'Impoverished Algorithms: Misguided Governments, Flawed Technologies, and Social Control' (2019) 46 Fordham Urban Law Journal 364.

[36] NITI Aayog, 'Approach Document for India Part 1 – Principles for Responsible AI' (February 2021) <https://www.niti.gov.in/sites/default/files/2021-02/Responsible-AI-22022021.pdf>.

[37] The Royal Society, 'Explainable AI: The Basics, Policy Briefing' (November 2019) accessed at <https://ec.europa.eu/futurium/en/system/files/ged/ai-and-interpretability-policy-briefing_creative_commons.pdf>; Yavar Bathaee, 'The Artificial Intelligence Black Box and the Failure of Intent and Causation' (2018) 31(2) Harvard Journal of Law & Technology 889.

[38] Soaad Qahhar Hossain and Syed Ishtiaque Ahmed, 'Towards a New Participatory Approach for Designing Artificial Intelligence and Data-Driven Technologies' (2021) accessed at <https://arxiv.org/ftp/arxiv/papers/2104/2104.04072.pdf>.

[39] Archon Fund, 'Varieties of Participation in Complex Governance' (2006) 75 Public Administration Review 66.

# PAI as a method of dealing with these problems | Benefits and Challenges

The origins of participatory design bear a striking resemblance to the current lack of involvement around the design and deployment of AI systems. Participatory design finds its origins in Scandinavian countries, commencing as a way for workers to be involved in the design process during an era of computerisation and digitisation in the 1970s.[40] There is a parallel to the current exclusion of ordinary citizens in the decisions taken to deploy AI systems.

## *The Benefits*

PAI has the potential to offer multiple benefits which take care of the problems identified in AI operationalisation and implementation. These problems, as have been identified above, include the following - the decision on when AI is operationalised is not taken after discussing its implications with the persons who will be affected by such a decision (the transparency and inclusion problem); the opaque nature of AI's operations (the black box problem); and the consequent automation bias. These problems also result in disempowering persons in their market relations or their relations with public bodies. A proper operationalisation of PAI (covered in the next section) can take care of these in the following ways-

**Empowering citizenry agency in AI decision-making**: In most literature espousing PAI governance, the foremost idea is to counter unilateral, top-down decision making about AI deployment.[41] For instance, in India, we have seen such application in sectors of education,[42] pension disbursement,[43] and especially in policing and law enforcement related activities.[44] In each instance, the lack of participation of relevant stakeholders, especially affected communities or specific sections of the population, has resulted in serious implementation challenges. For example, the Delhi Government's unilateral decision to install sophisticated surveillance systems in classrooms was met with severe resistance from both teachers and parents of the affected students. This has led to filing of writ petitions assailing this decision in the Delhi High Court.[45] PAI in theory and practice aims to bridge these gaps and pre-empt potential contentious breakdowns in implementation. For policymakers promoting the use of AI systems, engaging with stakeholders can facilitate a more enthusiastic adoption of such systems by the population. This is crucial in a country like India where adoption of technological solutions is significantly intertwined with behavioural perceptions and trustworthiness of any proposed interventions.[46] Hence, investing in building awareness and engagement is almost a *sine qua non* for successful implementation and scaling of any AI system solutions.

---

[40] Yngve Sundblad, 'UTOPIA: Participatory Design from Scandinavia to the World' in John Impagliazzo and others (eds.) History of Nordic Computing (Springer 2010).
[41] Iason Gabriel, 'Artificial Intelligence, Values, and Alignment' (2020) 30(3) Minds & Machines 411.
[42] Fareeda Iftikhar, 'Delhi: New school board to bank on AI for assessment' (Hindustan Times 21 March 2021) <https://www.hindustantimes.com/cities/delhi-news/new-school-board-to-bank-on-ai-for-tests-101617142057471.html> accessed 18 September 2023.
[43] 'Telangana government leveraging the power of AI and ML for pensioners' (IndiaAI, 27 October 2022) <https://indiaai.gov.in/case-study/telangana-government-leveraging-the-power-of-ai-and-ml-for-pensioners>.
[44] Vrinda Bhandari and Karan Lahiri, 'The Surveillance State, Privacy and Criminal Investigation in India: Possible Futures in a Post-Puttaswamy World' (2020) 3(2) University of Oxford Human Rights Journal 15; Pete Fussey and others, 'Assisted' Facial Recognition and the Reinvention of Suspicion and Discretion in Digital Policing' (2021) 61 British Journal of Criminology 325; Jai Vipra, The Use of Facial Recognition Technology for Policing in Delhi (Vidhi 2021).
[45] A reported copy of the petition is accessed at /https://images.assettype.com/barandbench/import/2019/05/CCTV-in-Delhi-schools-petition.pdf.

**Inclusion and Fairness**: A PAI framework is particularly crucial to unpack complex questions of bias and impact assessment of potential AI systems on the population as a whole, or sections of it. Particularly, while developers and deployers of AI systems may have incentives to endorse usage of a particular AI system, it is imperative that communities that are likely to be consumers, or in some capacity at the receiving end, also have an adequate understanding and say in such deployment. An example in India that demonstrates the dangers of non-participatory deployment are the use of FRT in policing and law enforcement. For instance, a study found that an increasing reliance on FRT and other predictive tools by Delhi Police, is likely to result in targeted policing against more vulnerable populations in the state.[47] Similar issues around exclusion due to lack of digital access or the general digital divide in India were also prevalent during the government back adoption of Aadhaar.[48] The positive impact of PAI is therefore, twofold - *first*, to prevent exclusionary deployment of an AI system, and *second*, where an AI system has been deployed and is likely to impact a particular community, to mitigate risks like bias, discriminatory output, and other potential harms.

**Ensuring critical feedback loops for *ex-post* evaluations and improvements**: A key part of PAI is the ability to share adequate and actionable feedback that can in turn aid with monitoring and evaluation of an AI system. Such feedback loops serve two main functions. *First*, in flagging functional glitches in an AI system for developers to address, and *second*, identification of some risks that may not have been evident during the *ex-ante* impact assessment. For example, when *Digiyatra* was introduced and piloted in 2022 at three airports, there were immediate posts on numerous social media platforms voicing technical glitches that some passengers faced.[49] However, there is little to no public information, definitely not in a published format, which may demonstrate if such *ad-hoc* feedback loops actually aided in the improvement of the concerned algorithm. Instead, as a subsequent paper of NITI Aayog has proposed, it would be ideal to have a more permanent and structured feedback loop for users and consumers particularly, to improve the overall functionality and accuracy of AI systems.[50] Similarly, public facing AI systems always have the potential to pose unforeseeable risks that become more evident through usage. Hence, feedback in this regard can aid in strengthening existing and future governance frameworks like laws, regulations, and oversight mechanisms.

---

[46] Ahmad Mushfiq Mobarak and Neela Saldhana, 'Remove barriers to technology adoption for people in poverty' (2022) 6 Nature Human Behaviour 480.
[47] Vipra (n 43).
[48] Swetha Totapally and others, State of Aadhaar: A People's Perspective (Dalberg 2020); 'Unseen and Unrecognized: The Indians Excluded from Aadhaar' (Haqdarshak 24 August 2023) <https://haqdarshak.com/2023/08/24/unseen-and-unrecognised-the-indians-excluded-from-aadhaar/#:~:text=Aadhaar%20identity%20among%20adults%20is%20considered%20%E2%80%9Cnear%20universal%E2%80%9D%20now.&text=However%2C%20studies%20have%20shown%20that,them%20without%20a%20recognised%20identity.> accessed 28 September 2023.
[49] A Shaji George and others, 'From Paperwork to Biometrics: Assessing the Digitization of Air Travel in India Through Digi Yatra' (2023) 1(4) Partners Universal International Innovation Journal 110.
[50] NITI Aayog, 'Adopting the Framework: A Use Case Approach on Facial Recognition Technology' (November 2022, Working Paper).

**Overall fostering of trust and bolstering adoption at scale:** The final, yet a vital benefit of PAI is to foster engagement within affected communities to ensure a sense of trust in the AI systems designed and deployed. Trustworthiness of AI systems has been a point of discussion for several years now.[51] While trustworthiness in design and development may be geared towards ensuring minimal false positives and false negatives, in AI governance, trustworthiness is more of a public perception barometer. To cultivate trust and confidence in the deployment of a potential AI system, it is vital to craft a meaningful participatory framework for citizens in how such systems are to be governed, including but not limited to developing oversight mechanisms and guardrails to safeguard fundamental (human) rights.[52] There is documented research that empirically fosters how a PAI governance framework can aid in fostering a positive public perception around an AI system, and even contribute to better adoption.[53]

## *The Challenges*

All the benefits highlighted - empowering citizens, building trust, bettering the quality of the final AI product - all deal, in some measure, with the transparency/inclusion problem and the black box problem. However, this is not to say that PAI itself does not have its own problems. Implementation of PAI suffers from multiple issues. Some of them are as follows-

**Co-optation**: Within any participatory framework, there is a risk of the movement's co-optation by a select few dominant actors with vested interests.[54] Co-optation typically happens to bolster commercial output, or to gain legitimacy. The latter, in particular, has been a strong trend in participatory movements, where vested interests have manipulated stakeholders to be considered as proponents of the movement. However, the motives often vary from those of promoting real participation and engagement. In the Indian context, co-optation is not an unknown sentiment. We have often witnessed claims of such situations especially in political rallies and movements. Within PAI too, co-optation by dominant interests is a real challenge. Not only do such tactics undermine the legitimacy of the participatory efforts, but in most cases the co-opting entities act in a manipulative manner which is counter-productive to the larger interests of participating stakeholders.[55] For instance, there are concerns voiced by several distinguished scholars against long-termist ideas of AI risks.[56] They argue that proponents of this philosophy often are tech magnates who want to use purported existential risks of AI as a smokescreen to prevent meaningful regulation at present.[57] This presents a classic scenario of co-optation where Big Tech backers are deemed as conscientious individuals working towards aiding AI regulation, but effectively creating a significant distraction for policymakers and legislators around existing risks of AI. In India, the risk of co-optation is also aggravated by the fact that the digital divide is severe. Its prevalence cuts across the urban-rural and educated-uneducated divides. In fact, many educated, urban citizens are likely to be unaware of tech implications, as has been demonstrated in several empirical studies around privacy perceptions of Indians.[58] In such a scenario, the potential for manipulation and co-optation is more probable.

---

[51] NITI Aayog (n 49), UNESCO's AI principles can be accessed here <https://www.unesco.org/en/artificial-intelligence/recommendation-ethics>
[52] Inclusive AI Governance (Ada Lovelace Institute 2023); Anton Sigfrids and others, 'Human-centricity in AI governance: A systemic approach' (2023) 6 Frontiers in Artificial Intelligence 1.
[53] Ada Lovelace Institute, 'How do People Feel About AI? A nationally representative survey of public attitudes to artificial intelligence in Britain' (Ada Lovelace Institute 2023) accessed at <https://www.adalovelaceinstitute.org/report/public-attitudes-ai/#executive-summary-2>.
[54] Sherry Arnstein, 'A Ladder of Citizen Participation' (1969) 35(3) Journal of American Institute of Planners 216.
[55] John P. Kotter and Leonard A. Schlesinger, 'Choosing Strategies for Change' (2008) Harvard Business Review <https://hbr.org/2008/07/choosing-strategies-for-change>. For more general discussions on co-optation in historical political scenarios, Abeba Birhane and others, 'Power to the People? Opportunities and Challenges for Participatory AI' (2022) EAAMO 1.
[56] 'Doomer AI advisor joins Musk's xAI, the 4th top research lab focused on AI apocalypse' (VentureBeat 24 July 2023) <https://venturebeat.com/ai/doomer-advisor-joins-musks-xai-the-4th-top-research-lab-focused-on-ai-apocalypse/>.
[57] See for eg, Sigal Samuel, 'Effective altruism's most controversial idea' (Vox 6 September 2022) <https://www.vox.com/future-perfect/23298870/effective-altruism-longtermism-will-macaskill-future>

**India's digital divide and lack of non-expert viewpoints in PAI governance**: A concern that is arguably accentuated in India is that of digital divide and limited digital literacy. While many researchers have commented on how India's digital divide impacts an individual's ability to access digital goods and services,[59] it is likely to also impact any potential participatory frameworks for AI governance. PAI is premised on the ability of stakeholders to meaningfully engage and contribute to oversight mechanisms within the AI ecosystem. This requires such stakeholders to have a developed understanding of how such systems are likely to be deployed, the risks they pose (both foreseeable and unanticipated) and propose innovative solutions for risk mitigation. In India, where even rudimentary digital literacy in general is limited,[60] it is not farfetched to assume that even a theoretically well-constructed participatory framework may arguably fail to bring a wide array of stakeholders because of lack of understanding of AI systems and their respective risks. In fact, existing forms of stakeholder consultations in India with respect to AI governance, have largely been limited to three main categories of stakeholders namely, industry representatives, bureaucrats driving legislation in this field, limited civil society participants (mostly legal researchers, and technology academics). To supplement these more experts-driven consultations, it is imperative for non-experts to also find ways to plug themselves into these conversations. General civil society interactions through online polls or surveys to solicit inputs, is an example but actual instances of such tactics are scarce to find in India. Given the dominance of expert opinions, and the lack of adequate methods for leveraging non-expert viewpoints, it is imperative that any potential PAI governance framework(s) address this deficiency.

**Participatory washing and tokenism**: At its core, like any participatory measure, PAI needs to be meaningful and effective and not merely a box checked for compliance.[61] Therefore, the steps that are proposed in the next section with respect to identification of relevant participants and their incorporation in decision making around AI systems, must be an exercise that displays robustness and inclusivity. While inclusivity itself is not the only feature of a participatory model, it is certainly one of the more important objectives. PAI governance frameworks conventionally seem to have limited the discourse to certain domain experts (like law and policy, technologists, public administration, and bureaucracy). Even relevant social scientists who can provide valuable empirical evidence to guide PAI, are often missing from the action.[62] What results is normally an echo-chamber phenomenon which converges on a few interests rather than evaluating a prospective AI use case in a more holistic manner. In India, we have seen (as discussed in Table 1), how some sectoral legislation and governance has managed to accomplish a more meaningful participatory approach. This needs to be replicated for AI governance as well, to ensure the representation of broader societal interests, and a more critical evaluation of risks posed by AI systems to more vulnerable (and less represented) sections of the populace.

---

[58] Abhishek Gupta and others, 'The Privacy Conundrum: An Empirical Examination of Barriers to Privacy Among Indian Social Media Users' in Sudhir Krishnaswamy and Divij Joshi, The Philosophy and Law of Information Regulation in India (CLPR 2022); Amol Kulkarni and others, 'Users' Perspectives on Privacy and Data Protection' (CUTS 2019).
[59] Basu Chandola, 'Exploring India's Digital Divide' (ORF 20 May 2022) <https://www.orfonline.org/expert-speak/exploring-indias-digital-divide/>.
[60] Amit Singh Khokhar, 'Digital Literacy: How Prepared is India to Embrace it?' (2016) 7(3) International Journal of Digital Literacy and Digital Competence 1; Pradipta Mukhopadhyay, 'A Case Study on Digital Literacy with Respect to India' (2021) 6(2) International Journal of Advanced Research in Science, Communication and Technology 1502.
[61] Ivana Bartoletti, 'Towards AI Transparency: Can a Participatory Approach Work?' (Medium, 21 October 2021) <https://medium.com/@reshaping_work/towards-ai-transparency-can-a-participatory-approach-work-c76f228bd01>.
[62] 'At discussions on AI ethics, you'd be hard-pressed to find anyone with a background in anthropology or sociology' (Times of India 14 January 2021) <https://timesofindia.indiatimes.com/blogs/the-interviews-blog/at-discussions-on-ai-ethics-youd-be-hard-pressed-to-find-anyone-with-a-background-in-anthropology-or-sociology/?source=app&frmapp=yes>.

**Transparency paradox and PAI:** With a growing recognition of inherent risks posed by AI systems, a key demand that has emerged is that of more transparency of AI models.[63] Even in India, NITI Aayog listed transparency and explainability as one of the seven principles of responsible AI application.[64] However, in response to more transparent AI models, a concern that has gained traction is the potential hacking and corruption of AI models by malicious actors. This phenomenon has been labelled as the AI transparency paradox.[65] Researchers have demonstrated that explanations and information shared about algorithms can make them susceptible to being reengineered or manipulated, and even the whole algorithm being stolen in violation of intellectual property rights of developers.[66] This transparency paradox can also occur in PAI, particularly during the design and development stages, where the objective is to devise oversight mechanisms based on an intricate understanding of the functioning of algorithms. Such an understanding can arguably be exploited by malicious actors who may embed themselves in the participatory process and gain vital details to exploit its vulnerabilities. Hence, it is imperative to determine what information can be shared, vetting participants adequately, and ensuring safeguards in place to pre-empt and prevent malicious actors.

The presence of both benefits and challenges means that the operationalisation of PAI must be granular, considering the perspective of all persons involved. However, determining who is 'involved', and how 'granular' such involvement must be, is itself a contested question. The next section attempts to concretise discussion on the topic.

---

[63] OECD, 'OECD AI Principles Overview' (May 2019) accessed at <https://oecd.ai/en/ai-principles>; NITI Aayog (n 49); UNESCO AI principles, accessed here <https://www.unesco.org/en/articles/recommendation-ethics-artificial-intelligence>; US AI Bill of Rights accessed here <https://www.whitehouse.gov/ostp/ai-bill-of-rights/>.
[64] NITI Aayog (n 49).
[65] Andrew Burt, 'The AI Transparency Paradox' (Harvard Business Review 13 December 2019) <https://hbr.org/2019/12/the-ai-transparency-paradox>.
[66] Smitha Milli and others, 'Model Reconstruction from Model Explanations' (2018) accessed here <1807.05185.pdf (arxiv.org)>.

# Operationalising PAI

## *Stages of AI Development*

Operationalising an AI based solution is always bound by practical considerations. Production grade AI implementation by any institution has plenty of opportunity for participation. This section hopes to provide an overview into the technical aspects of an AI based solution, and the possible avenues of participation. The AI system lifecycle phases are as follows: the design phase, the development phase and the deployment and adoption phase.[67]

### Design

The first phase ('Design') involves identifying and formulating a problem statement and reviewing relevant literature, and preparation, exploration, and sourcing of data. An approach is chosen depending on the problem statement and data available. The approach is also dictated by the quality, nature, and size of the data. Once the data is sourced, it might have inherent problems such as coherency, data corruption, and missing values. Additionally, gaining knowledge of overall trends is also helpful in making preliminary design choices. Thus, the data must be cleaned and explored. While the intricacies of these processes are explored in the following sections, the general idea is to check the integrity of the data and get an overall idea of what one is working on. The intuition behind this is that since much of the work on AI depends on the data, the quality of said data needs to be beyond question. The developer also needs a general idea of the structure so as to know what operations can be performed on the data and draw a boundary on the possibility of the experiments one can run. These steps impart direction to the development phase, helping develop relevant models and algorithms. The sourcing of data has various aspects to it. Data collection, collation, verification, annotation etc., are a few such aspects. Creation of datasets through these tasks becomes an integral part of the AI ecosystem.

Datasets are the lifeblood of any AI lifecycle and are constructed over various domains (like healthcare and law) and their inherent niches/ problem spaces (like cancer detection and risk assessment respectively) that might need an AI solution. For instance, if one needs an AI solution in the healthcare domain, one will need a comprehensive data of entities within that domain. Take the example of a potential tech solution that detects osteosarcoma; it would need scans of the afflicted long bones for training and scans that would give information on how the long bone should look if healthy. In this case, the 'entities' are the scans themselves.

### Development

The development phase involves building an initial model to measure a benchmark and trying different models and evaluating their primary metrics to achieve the best performance. Scholars break down the development phase into Task, Models and Features.[68] A 'Task' is a problem that can be solved by machine learning. It refers to the problem statement and its possible solution from an algorithmic standpoint. To gain any insight as to what algorithmic application might be a suitable fit one must look for cues from the structure of the data. This is an important step, since the structure of the data determines future decisions. For example, in the context of a spam classifier, the structure of the data can either be a simple yes/no classification- mails that are marked 'yes' will be sent to the spam folder, and those that are marked 'no' will not be. In this case, the structure of the data is a binary classification. However, it could also be along a scale of urgency, with the variable being continuous- that is, the variable is not a simple yes/no but is instead any number along a sliding scale. Mails that are not urgent at all (as determined by the programmer) will then be sent to the spam folder.

---

[67] Kevin Desouza and others, 'Designing, developing, and deploying artificial intelligence systems: Lessons from and for the public sector, Business Horizons' (2020) 63(2) Business Horizons 205-213; IT Modernization Centers of Excellence, 'Understanding and Managing the AI Lifecycle' accessed here <https://coe.gsa.gov/coe/ai-guide-for-government/understanding-managing-ai-lifecycle/>.
[68] Peter Flach, Machine Learning: The Art and Science of Algorithms that Make Sense of Data (Cambridge 2020).

Once a sense of the structure is attained, the next step is the Model stage. It can be viewed as the output of machine learning. Much akin to maths, there are various approaches one can take to solve a problem. They can be distinguished either on their main intuition, giving us approaches like geometric models, probabilistic models, logical models etc, (models where the decision is made on the basis of a geometric inference, a probability distribution, and answers to yes/no questions respectively). They can also be distinguished on their *modus operandi*, giving us approaches like 'Grouping' and 'Grading' models (grouping refers to the process of clustering similar data points together, while grading involves assigning labels or scores to data points based on certain criteria). Depending on what the modality of the required output is, a model can be chosen to work on the problem. The main aim of choosing a model is to choose an approach that intuitively and mathematically makes sense in the search for a solution. It reinforces the idea that machine learning, and in a much larger sense AI, is not a black box but a set of carefully crafted mathematical decisions.

The final ingredient to machine learning are the 'Features'. Data and a model to work on said data alone are not sufficient in terms of solving problems. There is a plethora of cases where there are allowances that need to be made due to the inherent characteristics of the data, data complexity or data gaps. This is where feature engineering comes into play. For example, in a problem with a high number of features, it would make sense to either eliminate the features that add very little value or combine all the features to form a new set of composite features that are fewer in number.

The choice of how this trio of Task, Models, Features is constructed is often driven by Metric and 'Benchmarking'. Intuitively, one would choose the combination that gives the best performance on the choice of metric. Metrics are essentially a means of measuring error and aiding the task of minimising it to improve performance. While this might seem like an intuitive way to go for the approach that gives the best performance, the notion of error itself is something that must be thought about. The nature of the task (ex: regression, classification etc.) is one such factor that determines the kind of error we measure and by extension the choice of metric. The cyclical process of optimiation on the basis of these appropriately chosen metrics is the foundation of the development phase and leads to one getting the best possible fit for a solution.

Understanding these facets of development are important as there are various decisions made throughout this phase. Understanding the importance and effects of each decision not only improves the decision-making process but, in the context of this paper, also gives us clarity on who should be involved in the decision-making process, how much power they should wield and how much information from other stages is relevant for making an informed choice. Answering these questions becomes integral to establishing a more participative structure in the creation of AI systems.

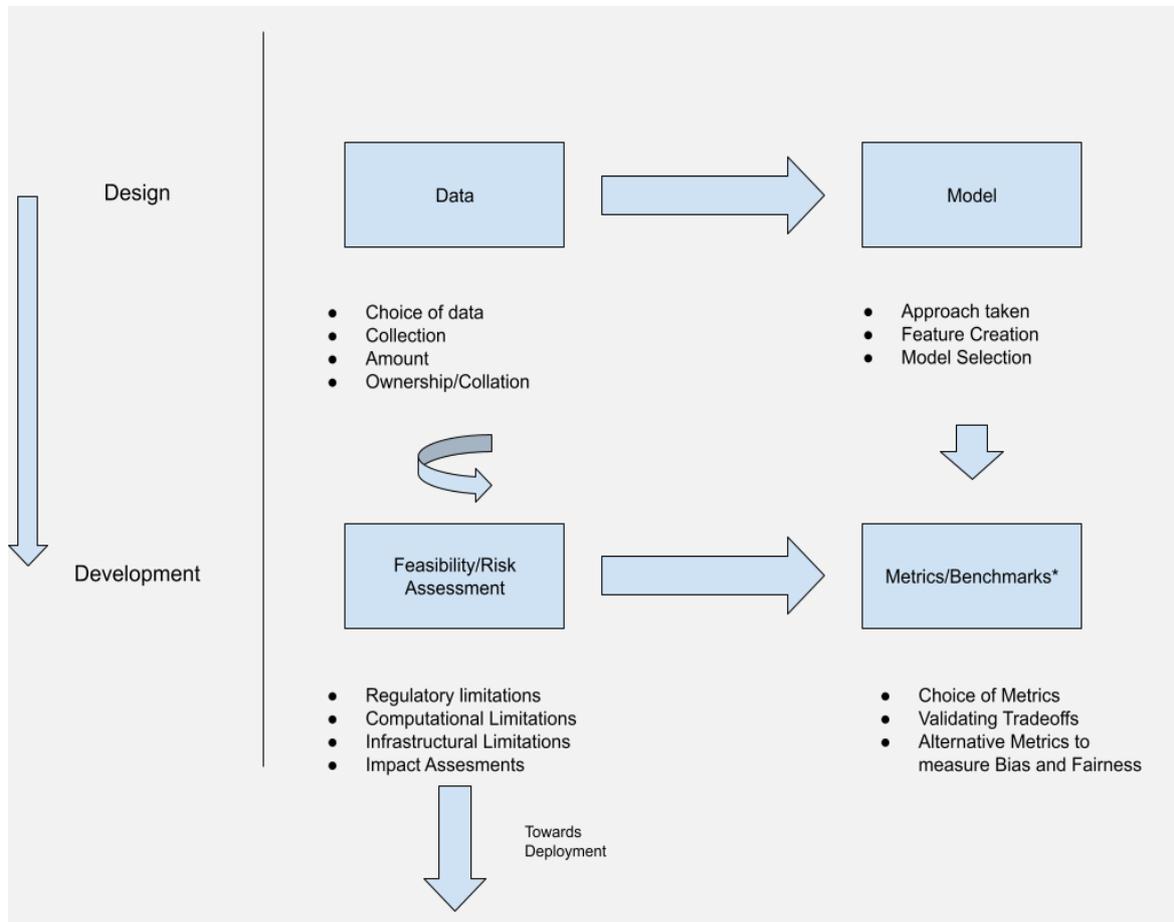

**Figure 1:** *The above diagram illustrates the various aspects of participation at different stages of the AI life cycle. While most of these aspects are self-explanatory or explained in detail in the previous sections, the various aspects of Metrics and Benchmarking provide an interesting landscape of decisions.*

We must acknowledge that real-world biases can creep into the solution. There has been quite a lot of published evidence of uneven treatment of different demographics,[69] and this can reportedly cause a plethora of serious problems. Some of these issues are unfair allocation of opportunities or unfavourable representation of particular social groups.[70] To ensure that every end user gets the promised performance independent of other extraneous factors, models must be vetted using bias and fairness metrics. Czarnowska and others,[71] go into detail in describing various fairness metrics, and introduce three of their own scoring metrics, each conceptually different to the other. The choice of the metrics depends on many factors, including the task, the particulars of how and where the system is deployed, as well as the goals to be accomplished. Thus, they recommend that the choice of metric be grounded in the application domain, enabling one to choose a few but relevant and appropriate metrics to measure the performance of one's models.

---

[69] Lucas Dixon and others, 'Measuring and Mitigating Unintended Bias in Text Classification' (2018) Proceedings of the 2018 AAAI/ACM Conference on AI, Ethics, and Society 67; Daniel Borkan and others, 'Nuanced Metrics for Measuring Unintended Bias with Real Data for Text Classification' (2019) Companion Proceedings of The 2019 World Wide Web Conference 491-500; Nikita Nangia and others, 'CrowS-Pairs: A Challenge Dataset for Measuring Social Biases in Masked Language Models' (2020) Proceedings of the 2020 Conference on Empirical Methods in Natural Language Processing 1953-1967.

[70] Su Lin Blodgett and others, 'Language (Technology) is Power: A Critical Survey of "Bias" in NLP' (2020) Proceedings of the 58th Annual Meeting of the Association for Computational Linguistics 5454–5476.

While these are steps taken to get the ball rolling, production grade AI needs measures to keep the ball rolling. In the world of AI/ML) there is a concentrated effort on designing and developing models, which needs to be assisted by deployment and post deployment measures to ensure these models reach production. There are various issues that might arise once a solution is deployed. Ranging from inefficient workflows and bottlenecks to changing domain regulations and model drifts (the decay of models' predictive power as a result of the changes in real world environments),[72] there are a plethora of problems that need attention so as to ensure performance. To ensure that these problems are dealt with in the most efficient manner, solutions are usually developed with hooks built into them that help monitor, edit, and improve solutions. These sets of practices that detail how to roll out machine learning models, monitor them, and retrain them in a structured and segmented manner, are called 'MLOps'. MLOps, as defined by Kreuzbergeret al,[73] is a paradigm, including aspects like best practices, sets of concepts, as well as a development culture when it comes to the end-to-end conceptualisation, implementation, monitoring, deployment, and scalability of machine learning products. Essentially, MLOps aims to facilitate creation of AI/ML products while ensuring the existence of principles like automated integration and deployment of new features; reproducibility of the solution; versioning of the data, model, and code; continuous training and evaluation of the models; continuous monitoring; and feedback loops. MLOps acts as a structural solution to adapt to any real-world challenges that might appear. Problems like changing regulations, change in policy, change in data quality, change in the demographic the solution is used on, newly found adverse effects, etc, might need measures like retraining the model, reworking the concept and deploying the new version/features, or even withdrawing the solution. In such situations, the MLOps paradigm provides a steady base to be able perform any of these actions in the best possible manner.

---

[71] Paula Czarnowska and others, 'Quantifying Social Biases in NLP: A Generalization and Empirical Comparison of Extrinsic Fairness Metrics' (2021) 9 Transactions of the Association for Computational Linguistics 1.
[72] 'What is Model Drift?' (Dominoe AI) <https://domino.ai/data-science-dictionary/model-drift> accessed 16 January 2024.
[73] Dominik Kreuzberger and others, 'Machine Learning Operations (MLOps): Overview, Definition, and Architecture' (2022) 11 IEEE Access 31866-31879.

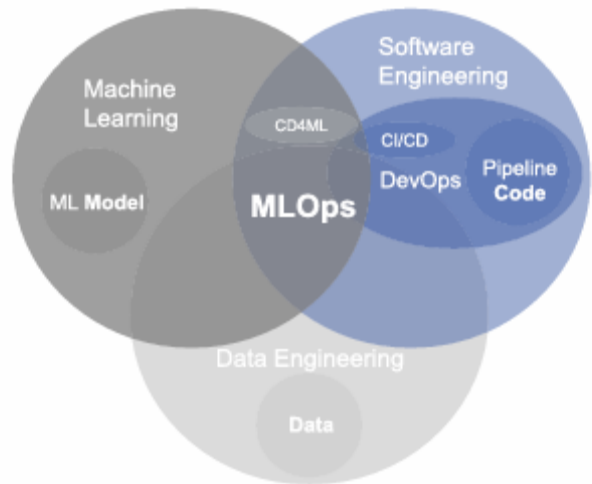

**Figure 2:**[74]*The given figure is a venn diagram that shows the relations of MLOps with other well defined disciplines.*

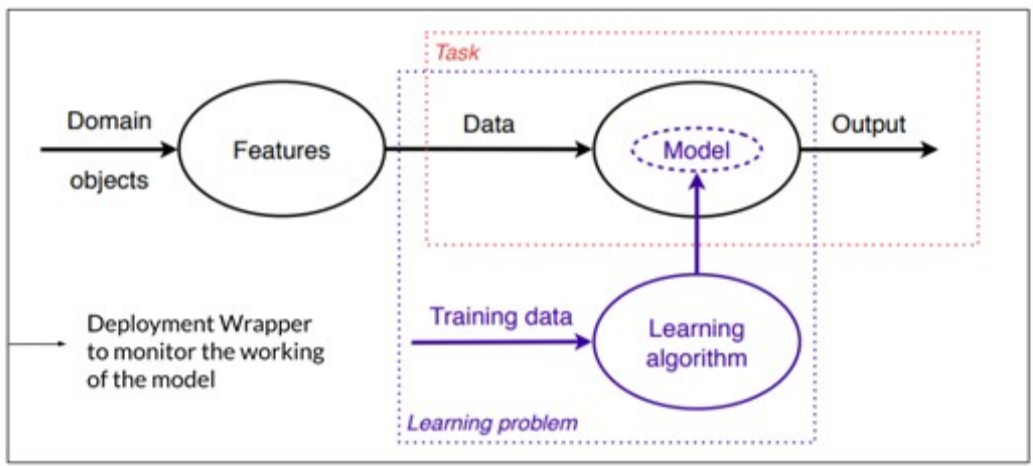

**Figure 3:**[75] *An overview of how machine learning is used to address a given task. A task (red box) requires an appropriate mapping- a model- from data described by features to outputs. Obtaining such a mapping from training data is what constitutes a learning problem (blue box).*

## Deployment and Adoption

Once the AI model has been developed, deployment - putting the model into actual use- begins. This requires the transformation of an AI model or prototype into an integrated solution within a specific ecosystem (say a predictive algorithm determining the credit score of a potential loan applicant).[76] Deployment is an overarching term for different sub-stages. It comprises initial implementation and risk assessment of an AI model, effectuating governance frameworks through legislation, regulations and standard setting, creating ex-post monitoring and evaluation frameworks, and establishing feedback loops from end users/deployers, and third-party affected persons.[77]

---

[74] ibid 31879.
[75] Flach n (67) 11. Necessary additions have been made by CeRAI.
[76] Daswin DeSilva and Damminda Alahakoon, 'An artificial intelligence life cycle: From conception to production' (2022) 6(3) Patterns 1.

Initial implementation of an AI model predominantly focuses on risk assessment. As AI systems are increasingly being adopted in complex real-world situations, it is necessary that a comprehensive risk assessment be conducted for such scenarios. In this regard, *ex-ante* impact assessments, risk audits,[78] and even more specific assessments like the danger to fundamental (human) rights,[79] have become mainstream. The risk assessments cover actual or potential problems that the model may possess regarding privacy, cybersecurity (or susceptibility to hacking), trustworthiness, explainability of the model, robustness, usability, and its social-technical implications (especially on vulnerable populations).[80] The same needs to be systematically catalogued into a risk register and be addressed through effective governance.

The governance sub-stage is crucial from a risk mitigation perspective. AI governance has been a key discourse in larger conversations of responsible adoption of AI systems. At present, there are multiple indices which rank how AI governance is being undertaken in different countries.[81] Whether through formal modes of legislation, or less hardcoded models like standard setting guidelines, self-regulatory models, etc., this part of deployment focuses on establishing guardrails. The ideas of PAI based governance derive significantly from social science literature (discussed above) on how regulation and oversight can be made an empowering exercise rather than a tokenistic one (or participatory washing).[82] For AI governance, two key aspects are setting out liability standards for actual harm caused by an AI system and establishing effective measures for continuous oversight and grievance redressal. Besides this, governance may also address ancillary issues like impact assessments, monitoring and evaluation and feedback formats, as discussed next.

Monitoring and evaluation essentially allow the AI model to evolve based on user preferences and changes in real world scenarios.[83] Conducting periodic *ex-post* tech audits are being advocated as an active measure for monitoring and evaluation once an AI system has been deployed. Also termed as post-market monitoring, the process is crucial to ensure that high-risk AI models continue to conform with safety standards and not evolve in manners that become problematic. An example of monitoring and evaluation is present in the EU's proposed AI Act.[84] Any provider of an AI system is required to monitor the same once it has been deployed and

---

[77] ibid.
[78] For Humanity, 'Auditing AI and Autonomous Systems' (For Humanity) accessed at <https://forhumanity.center/article/auditing-ai-and-autonomous-systems-building-an-infrastructureoftrust/>.
[79] The EU AI Act recent text that was agreed upon at the end of the Trilogue is accessible at https://www.europarl.europa.eu/news/en/press-room/20231206IPR15699/artificial-intelligence-act-deal-on-comprehensive-rules-for-trustworthy-ai.
[80] For Humanity (n 77).
[81] As examples, see the Government AI Readiness Index accessed at <https://oxfordinsights.com/ai-readiness/ai-readiness-index/>; HAI, 'Artificial Intelligence Index Report' (2023) <https://aiindex.stanford.edu/wp-content/uploads/2023/04/HAI_AI-Index-Report_2023.pdf> accessed 16 January 2024.
[82] Renée Sieber and Ana Brandusescu, 'Final Report: Civic empowerment in the development and deployment of AI systems' (2021) accessed at <https://papers.ssrn.com/sol3/papers.cfm?abstract_id=4104593> .
[83] Jeff Saltz, 'What is the AI Life Cycle?' (Data Science Process Alliance 6 October 2023) <https://www.datascience-pm.com/ai-lifecycle/> accessed 16 January 2023.
[84] Artificial Intelligence Act 2023 (EU).

adopted, report any serious incidents or malfunctioning, and take corrective measures for the same.[85]

To complement the developer and deployer's monitoring and evaluation, it is also important to establish clear feedback loops. Feedback loops are a key step to ensure PAI in the deployment and adoption of an AI system. Having robust mechanisms for collating and processing public feedback is also crucial to establish trust in the AI system, which is a universally accepted principle of deploying responsible AI.[86]

The interplay between these three streams/phases of Design, Development and Deployment (and by extension post deployment) is the composite path to building a production grade AI solution that can be implemented by an institution. While these processes might seem discrete and are delved into separately, they do not exist in silos and need to be interconnected with each other. Inputs at any of these phases can be used in other phases as well.

From a PAI perspective, it is crucial to identify the relevant stakeholders for each of these phases of the AI life cycle, and distribute them appropriately to optimise the quality of inputs received. This is discussed in the next section.

## *Choice and Informational Access*

The technical overview of the processes involved in the creation of a solution, while shedding some light for a layperson, give rise to a series of questions, all with the overarching theme of choice. Now that one knows of the different phases, how does one make a choice at each of these stages? This section hopes to explore these questions and the general notion of choice and participation in the creation pipeline of a solution.

As Barbara Grosz,[87] Christopher H. Gyldenkærne and others have noted,[88] there are many problems that arise in AI when proper participation is not ensured. To that end, there must be space for participation through all the phases and a flow of relevant information to make decisions. The information that is required needs to be in a form consumable by the participant and must also be relevant. For example: the end user might not understand nor be able to participate fruitfully if the information available to them is the F1 Score metric, or just the different possibilities of models that can be used. The legal team might need information on the data collection and the model metrics in a manner that is consumable by them to ensure adherence to legal standards. One won't be able to derive fruitful participation from these groups without some form of pruning and translation of the information. It must also be noted that the transfer of information takes place in two planes, vertical and horizontal. Vertical transfer of information involves the transfer of information between two phases (of the AI lifecycle) and the horizontal transfer of information involves transfer of information within a phase. For example: the choices of models and metrics to benchmark involves a horizontal transfer of information to aid the decision-making process, whereas conveying the final decision to the next stage is a vertical transfer of information. While transparency and communicability are important aspects of horizontal translation, the interpretability and explainability of the solution are important aspects of the vertical translation. This is so because, as we will see, stakeholders are often optimally classified as per the stage of the AI development.

---

[85] Jakob Mökander and others, 'Conformity Assessments and Post-market Monitoring: A Guide to the Role of Auditing in the Proposed European AI Regulation' (2022) 32 Minds and Machines 241; A summary of the same is at Benjamin Cedric Larsen, 'Conformity Assessments and Post-market Monitoring: A Guide to the Role of Auditing in the Proposed European AI Regulation' (MAIEI 2 March 2022) <https://montrealethics.ai/conformity-assessments-and-post-market-monitoring-a-guide-to-the-role-of-auditing-in-the-proposed-european-ai-regulation/> accessed 6 January 2023.
[86] NITI Aayog, Adopting the Framework: A Use Case Approach on Facial Recognition Technology (November 2022, Working Paper); NITI Aayog, 'Approach Document for India Part 1 – Principles for Responsible AI' (February 2021) <https://www.niti.gov.in/sites/default/files/2021-02/Responsible-AI-22022021.pdf>
[87] Barbara Grosz, 'The AI Revolution Needs Expertise in People, Publics, and Societies' (Harvard Data Science Review 2 July 2019) <https://hdsr.mitpress.mit.edu/pub/wiq01ru6/release/5> accessed on 16 January 2024.
[88] Christopher Gyldenkaerne and others, 'PD and The Challenge of AI in Health-Care' (2020) 2 Proceedings of the 16th Participatory Design Conference 26–29.

These measures must be taken to ensure the efficacy and integrity of participation. Everyone involved need not be privy to all the information generated through all the processes but they must be privy to information relevant to them (this point is elaborated upon in the subsequent sections). While the identification of relevant groups who must be involved in the decision-making process and the hierarchy of information flow is discussed in later sections, it must be noted that none of these groups are limited to any one phase and can contribute horizontally or vertically depending on the decision at hand.

Thus, this paper proposes a sieve-like plane to visualise the flow of information and decisions. The identified stakeholders all exist on this structure we are calling the 'decision sieve'. The hierarchy of groups is decided depending on the context of a decision/choice that needs to be made and can be incorporated during the collation of choices/views which is dealt with in much greater depth in later sections. Here, the operational decision made is passed with relevant information translated as per the needs of the different groups. Decisions/choices from different phases can be passed through the sieve to obtain insights and inputs from relevant groups. The idea is to foster participation and ensure that the output of the sieve gives us a well-represented view of the different groups. This also helps control the flow of information, and control the groups involved per decision. Thus, the decision sieve aims to give a clearer view of the interplay of different groups making the plethora of decisions that an AI solution requires while ensuring that the decisions thus made are a product of participation. We believe this improves the quality of the solution and also helps gauge the impact of the solution. Find below an example of a decision sieve, with different groups identified and involved in various decisions.

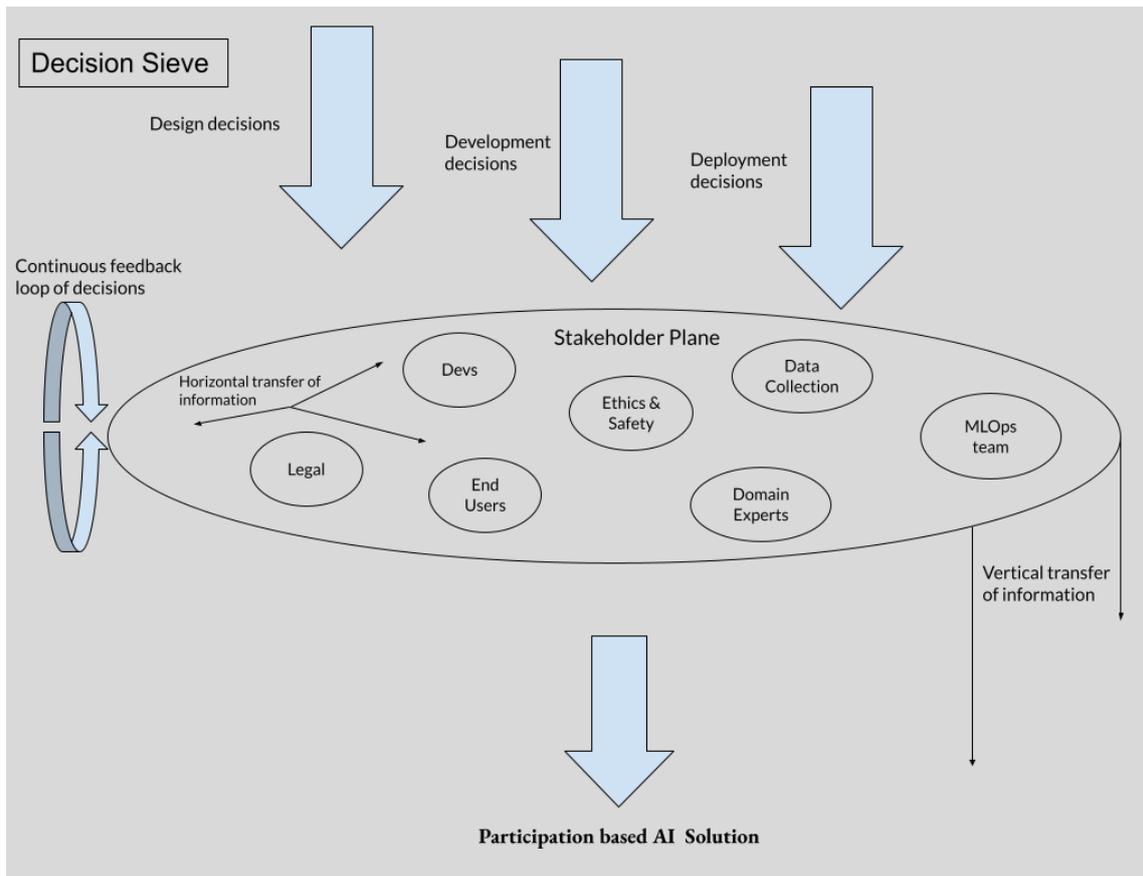

**Figure 4:** *Figure 4 represents an ideated 'decision sieve' through which all decisions requiring public participation ought to be passed before they are finalised. The arrows at the top represent the three types of decisions- design, development, and deployment. For each decision, the 'stakeholder plane' is determined according to the criteria of harm, urgency, legitimacy, and power. Information is shared among the participants as part of the participatory process.*

## Identification of Stakeholders

This part of the paper had started out with two questions- the question of how granular one must be while discussing PAI and the question of who counts as a stakeholder. The first question has been answered in the preceding section. The second question is sought to be answered here. Any discussion of participatory governance is futile if the question of who is chosen for the participatory exercise is not decided. Further, considering the variety of usages in which AI can be deployed, it also becomes important to provide a broad picture of who stakeholders are and what their roles can be. At the same time, overgeneralisations and overly vague answers also do not provide guidance. With these preliminary points, this section of the paper lays out the following – (a) who the stakeholders are; (b) the method by which they are identified; (c) what their level of involvement is (and ought to be), and (d) what stage they ought to be included in.

Scholars give four criteria by which stakeholders can be identified- power, legitimacy, urgency, and harm.[89] The first three of these criteria are borrowed from previous scholarship, which itself is based on an extensive literature review on stakeholder theory.[90]

*Power* is defined as the ability of one entity to impose its will on another. *Legitimacy* refers to a form of influence that seems to be justifiable even by those whose behaviour is affected. *Urgency* is how urgently a stakeholder's claims must be addressed. It can be either time-sensitive or relationship-sensitive. Finally, *harm* is how much a stakeholder's interests can be negatively affected by the implementation of the project.[91] For Mitchell, an entity that possesses two or more of the attributes is an important and salient stakeholder.[92] For instance, if the proposed project concerns the deployment of AI technologies in judicial settings (such as deciding on the future propensities of undertrials to commit crimes, known as 'risk assessments'),[93] then civil society groups which are capable of representing the interests of undertrial prisoners will be important. Undertrials are both capable of suffering direct *harm* from the application of the problematic autonomous decision-making software (**ADMS**), and their claim is also *urgent*. Similarly, judicial officers who are tasked with implementing these ADMS are also important stakeholders. They have a *legitimate* claim, in the sense that their decision on this issue is supposed to be final. Experts in algorithmic prediction will enjoy a similar position – they are legitimate because they are best able to testify as to the validity of the output that the algorithm produces. Government officials and politicians have power to influence the issue and may also be legitimate proxies for marginalised groups.

---

[89] Gloria Miller, 'Stakeholder Roles in artificial intelligence projects' (2022) 3 Project Leadership and Society 1.
[90] Ronald Mitchell and others, 'Towards a Theory of Stakeholder Identification and Salience: Defining Who and What Really Counts' (1997) 22(4) The Academic of Management Review 853.
[91] Miller further classifies harms, losses, and damages as follows: harms: bodily harm, loss of life, limitation of rights (freedom of movement), surveillance, psychological; loss: violations of human or civil rights such as loss of privacy, security, freedom, financial, job; damage: trust, reputation, environment. See Miller (n 88) 7. This classification is based on a prior study.
[92] Miller (n 88) 7.
[93] An example of this is the COMPAS algorithm, developed by Northpointe (private entity) and used by courts across the United States. See Julia Angwin and others, 'Machine Bias' (ProPublica 23 May 2016) <https://www.propublica.org/article/machine-bias-risk-assessments-in-criminal-sentencing>. Their empirical analysis revealed that black people were much more likely to be incorrectly labelled as high risk, and white people were much less likely to do so. In other words, the discriminatory ability of the algorithm was low, which was most likely a result of biased data due to structural racism. COMPAS is a proprietary software, which also means that it is not possible to know the weightage that is given to different factors. See David Spiegelhalter, The Art of Statistics: Learning from Data (Pelican 2019).

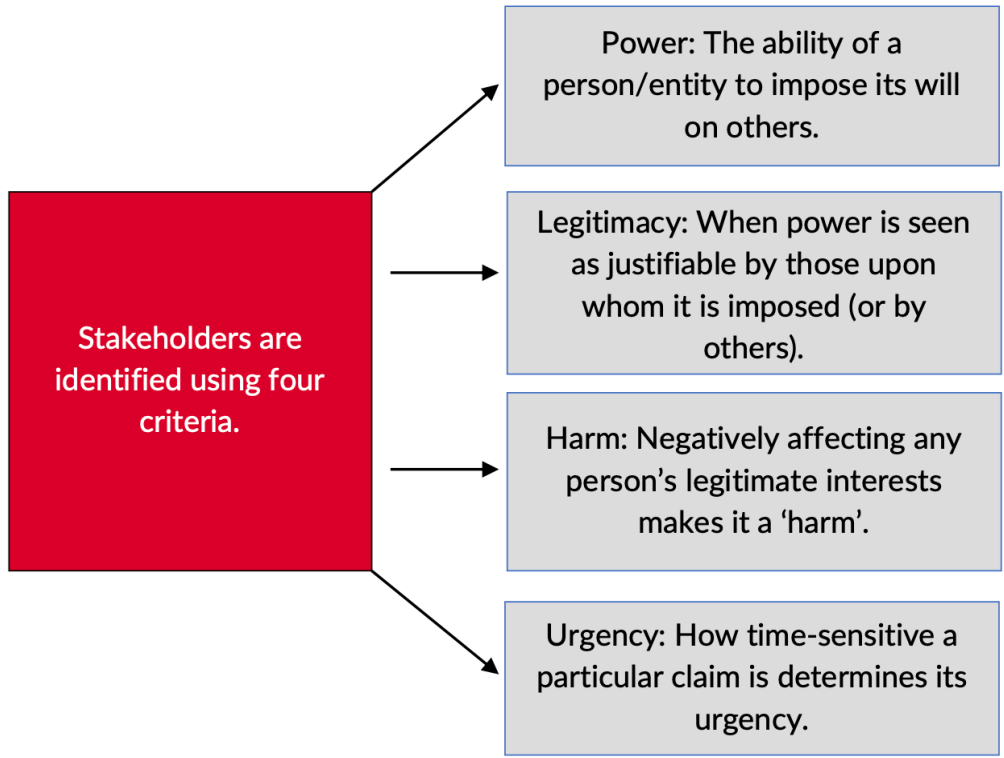

**Figure 5:** *Figure 5 lists out in a diagrammatic form the four criteria by which stakeholders are identified.*

As another example, in the Indian context, the *National Strategy for Artificial Intelligence* released by NITI Aayog points to various use cases in healthcare where AI has the potential to be transformative.[94] One of the examples which they point to is a machine learning solution to plug the gap created by a low supply of oncology specialists and a large number of cancer patients.[95] The database will consist of features which, according to the strategy paper, will enable even a general pathologist in making quality diagnoses on cancer.[96] Taking this example, there are multiple stakeholder who would be invaluable participants. Here, doctors and data scientists will be important stakeholders since they both have power (to administer and monitor such an algorithm) as well as legitimacy (due to their expertise). The general pathologists who will be responsible for execution are also legitimate and powerful stakeholders. Additionally, cancer patients who will be subjected to such an algorithm, and potentially stand to be affected through the use of such diagnostic or prescriptive algorithms (and potentially suffer life-threatening harm with inaccurate outputs), should also be included.

For any AI that is sought to be used by the public sector, government departments, implementing/nodal agencies will also be important stakeholders by virtue of their executive capacity.

---

[94] NITI Aayog, National Strategy for Artificial Intelligence (2018) <https://niti.gov.in/sites/default/files/2019-01/NationalStrategy-for-AI-Discussion-Paper.pdf>.
[95] ibid 28-29.
[96] ibid.

Two points must be noted here. First, the categorisations made above as to who is a legitimate, powerful, or an urgent stakeholder are all subjective classifications.[97] As far as possible, reasons have been given justifying an entity's categorization, although this is unlikely to matter in the final analysis as long as someone is included as a stakeholder. Second, another way to conceptualise stakeholders is to divide them along lines of developers, theorists, ethicists, and users (and affected persons).[98] Close examination will reveal that these categories map onto the categories developed by Miller and Mitchell. However, the Miller/Mitchell categorisation is more precise, and will be utilised throughout the rest of the paper.

At the end of the day, the concept of a stakeholder itself is a classification. Any classification is exclusionary if it is to be specific. If this is so, then the best that can be hoped for is to create criteria (heuristics) which are seen as both legitimate and implementable. Inherent subjectivity coupled with the plain fact that project managers (or in the case of AI in the public sector, any government department) have authority to decide these affairs should make one cautious about providing criteria that are universally applicable.[99] Lastly, to bring some certainty, formal statutory instruments can codify and define who can be a legitimate stakeholder in an AI system's design, development, and deployment.

## *Stakeholder Distribution*

This part of the paper starts with the premise that it is more feasible and implementable to map stakeholders as per their area of expertise/contribution to specific stages of the AI life cycle rather than to provide full discretion to program developers. This mapping need not be exhaustive, and exceptions will exist. However, insofar as it serves as a heuristic to channel participatory inputs optimally (i.e., where the inputs are most needed and are heard at the correct time), such mapping serves an important purpose in the overall discussion on PAI.

This is a slightly contentious issue as there are arguments to the contrary as well. Some argue that participation cannot be slotted thus, and that participation from *all* parties concerned may be required at all stages.[100] There is some merit in this point. Deciding, before information is revealed by stakeholders, that a certain stakeholder has information relevant only to a certain stage, is likely going to be an imperfect exercise. On a practical level, decision makers might fail to take inputs at the relevant time. To avoid these pitfalls, the argument stipulates the rationale for keeping the floor open to persons who wish to provide input at their desired time.

---

[97] Fernando Delgado and others, 'Stakeholder Identification and AI: Beyond "Add Diverse Stakeholders and Stir"' (2021) 35th Conference on Neural Information Processing Systems 1. "Despite their intentions to empower stakeholders and democratize the AI design process, those who "own" the participatory AI project had unparalleled authority in making decisions about participatory approaches in practice."
[98] "Affected Persons" is being argued by researchers to be an additional classification of people susceptible to AI harm, separate from users, as in many cases the two categories do not overlap. See, Connor Dunlop, 'An EU AI Act that works for people and society' (Ada Lovelace Institute 2023) accessed at <https://www.adalovelaceinstitute.org/policy-briefing/eu-ai-act-trilogues/#:~:text=5.,Protection%20and%20representation%20for%20affected%20persons,proposed%20by%20the%20European%20Parliament.>; Alun Preece and others, 'Stakeholders in Explainable AI' accessed at <https://orca.cardiff.ac.uk/id/eprint/116031/1/Stakeholders_in_Explainable_AI.pdf>.
[99] However, this is not to say that there have not been any attempts in this regard. See Batya Friedman and Peter Kahn Jr., 'Human Values, Ethics, and Design' in Andrew Sears and Julie Jacko (eds.) The Human Computer Interaction Handbook (Taylor and Francis 2008).
[100] This input was received in the consultation that Vidhi and CeRAI had organised for discussing this paper.

However, the error of this conclusion lies in mistaking mapping as *compulsion*, instead of a *guidance*. It is a fact that decision makers have discretion in deciding many things in the development of an AI algorithm (including who counts as a stakeholder, as previously discussed). The fact that some guidance is sought to be given does not mean that contrary action is not possible. Based on the specific facts, it may even be desirable. The discussion subsequent must be considered with this flexibility in mind and that the points therein can be deviated from (and adhered to where it is pragmatic).

Before delving further into stakeholder distribution, it is pertinent to mention the absence of this issue in most extant literature on PAI, as well as participatory governance in general.[101] Where it is mentioned, it is done so cursorily. For example, in the context of the Finnish public sector, one respondent in a study stated that participation ought to be maximised at the start and the end of the entire process.[102] As such, a reference to mapping and distribution more generally becomes necessary.

A form of mapping can sometimes be conducted in an 'hourglass model'. For example, political scientists argue, in the context of constitution making processes, that participation ought to take place in the initial stages and in the latter stages; the period in the middle should best be piloted by a closed assembly capable of deliberation and compromise.[103] The 'hourglass' model takes its significance from the type of project for which consultation is required. The framing of a constitution takes place under historically contingent circumstances, and requires long-term thinking based on commonly shared principles. Elster's point is that the hourglass model, where participation is maximised in the initial and final stages of the constitution making process (the initial part can include voting on the type of assembly, deciding the terms of reference of the assembly, if any, etc. and the latter part will include ratification, debate, etc.) will create an environment that is best suited for deliberation and compromise.[104] These results are obviously not directly portable into the context of PAI. Nevertheless, a similarity exists insofar as both are projects attempting to balance expertise and broad-based participation. The dangers of participatory whitewashing, co-optation, and others also exist in the constitution making context as well. The analogy ought not to be pushed too far. Its only value lies in conceptualising the hourglass model as a possible method of participatory classification independent of the process to which it is applied. Beyond that, the scale of the two projects is too different to make any accurate comparisons.

---

[101] A substantial amount of literature on this point is very silent on this topic. As examples, see Vincent Luyet and others, 'A framework to implement stakeholder participation in environmental projects' (2012) 111 Journal of Environmental Management 213; Ciara Fitzgerald and others, 'Citizen participation in decision-making: can one make a difference?' (2016) 25(1) Journal of Decision Systems 248; Archon Fung, 'Varieties of Participation in Complex Governance' (2006) 75 Public Administration Review 66; Elizabeth Bondi and others, 'Envisioning Communities: A Participatory Approach Towards AI for Social Good' (2021) Proceedings of the 2021 AAAI/ACM Conference on AI, Ethics, and Society 425; Steven Umbrello, 'Mapping value sensitive design onto AI for social good principles' (2021) 1 AI and Ethics 283; Paola Sabina Lupo Stanghellini and Dennis Collentine, 'Stakeholder discourse and water management– implementation of the participatory model CATCH in a Northern Italian alpine sub-catchment' (2008) 12 Hydrology and Earth System Sciences 317; Henning Sten Hansen Milla Mäenpää, 'An overview of the challenges for public participation in river basin management and planning' (2008) 19(1) Management of Environmental Quality 67; Helen Sharp and others, 'Responsible AI Systems: Who are the stakeholders' (2022) Proceedings of the 22 AAAI/ICM Conference on AI, Ethics, and Society 227.

[102] The exact quote, as per the study, is as follows: "The early development phase, where you come up with the ideas and solutions and try to understand what would be the solution that works, that part they should be involved. And then, once that is implemented, and in use, they feel empowered, and then they have the agency to change that". Karolina Drobotowicz and others, 'Practitioners' perspectives on inclusion and civic empowerment in Finnish public sector AI' (2023) Proceedings of the 11th International Conference on Communities and Technologies 108, 114.

[103] Jon Elster, 'Legislatures as Constituent Assemblies' in Richard Bauman and Tsvi Kahana (eds.) The Least Examined Branch: The Role of Legislatures in the Constitutional State (Cambridge 2006) 197.

[104] More specifically, his argument is that an hourglass model in a constituent assembly convention (that is, an assembly which does not function as a parallel legislature) will reduce the influence of passion and interests and promote genuine compromise and deliberation. These considerations are, of course, very different in the context of an AI project. Here, the need for secrecy is not to reduce the influence of passions and interests (although that is also possible). Instead, taking into account the fact that a PAI approach ought to be utilized in as many cases as possible, this mapping attempts to make the same implementable by agencies which do not have expertise and the necessary wherewithal to conduct participation at every stage. See Elster (n 102) 196.

With this caution in mind, the first point for consideration in mapping stakeholder distribution, is of specialisation/competence/information of the participant. This must decide what stage of the process they are included in. For instance, including non-specialists in the development stage could result in the project managers having to spend a considerable amount of time in training them about the technical details to then be able to elicit feedback from them. While this is possible, the costs, both monetary and in terms of the time expended, may outweigh the benefits. An optimal solution may thus map specialists/experts to those stages which require specialised input. Non-experts are ordinarily better suited to the initial and final stages of the overall development. The initial design stage is when the boundaries of the project are decided via the problem statement, available literature, quality of the data, etc. The design of the problem statement, i.e., questions on whether the problem really requires a technical solution, and if yes, the type of such a solution, is where inputs from a wide variety of stakeholders can be aptly utilised. In the context of the decision sieve, this means that for each stage (representing the vertical arrow, bound in the stakeholder plane), the 'sieve' of stakeholders will be different based on factors that are highlighted below. Their input on the information that is given to them will then be used for subsequent stages. For example, in the case of a risk-assessment software being used in the judiciary, these discussions would centre on whether a risk-assessment software would discriminate or yield a biased output against certain individuals. Non-experts who have an urgent/legitimate need may be able to voice their concerns about the impact of the AI algorithm on their lives and interests. This will then inform the formulation of the problem statement itself, and indirectly affect even the technical aspects of developing the AI solution. Development, as we have seen, is a highly technical process. The task/model/features distinction in the development stage requires input from data scientists, software engineers, and other experts. Hence in most cases, it ought to be left to technical experts.

A second point to note, in this context, is that the process must consider the urgency of various stakeholders. The mapping of stakeholders to stages must not only be as per their expertise, but also their urgency. Some claims are time-bound and will have to be resolved as soon as the developers are made aware of them. For instance, if non-experts have crucial inputs on the manner in which the feature engineering process is being carried out, then that must take precedence over the general hourglass model stated above. This is especially important because the feature engineering stage, where raw data is transformed into a machine-readable form, is where many biases will creep into the dataset, affecting predictions and harming persons.

Expertise and urgency hence feature as important considerations in deciding when exactly a stakeholder ought to contribute. As mentioned before, these are not strict criteria that apply in a yes/no fashion, but rather act as broad reasons which justify the inclusion/exclusion of a stakeholder from a particular stage of the development process.

Finally, it is important to note that powerful stakeholders will always be able to find their way into the participatory process. Insofar as that is the case, the criteria for deciding stakeholder classification are the same as the criteria for deciding who is a stakeholder in the first place. Unlike the section on stakeholder identification, however, it is possible to hierarchise the three criteria. For instance, the primary criteria for deciding the stage where someone ought to be included is, under ordinary circumstances, urgency. This is so by the very nature of that criterion. Not addressing an urgent claim will risk losing time-valuable information. If this information is useful enough, then it may involve changes to the direction of the project, and course-correction may not be possible later. The second criteria, normatively speaking, is that of expertise. Keeping the hourglass model in mind, the participation of non-experts should be limited in the middle of the process, where the actual AI algorithm is being fleshed out. Power, normatively speaking, should be the last criteria. If a powerful stakeholder has either urgency or expertise, they should be prioritised *on that count,* and not because of their power. However, this schema is very unlikely to play out, since powerful stakeholders will have more control over the overall process.

These criteria then decide who is in the 'sieve', and the information that is given to them in order to elicit their inputs. However, whether inputs are elicited *before* a course of action is decided upon, or after the same, is up to the discretion of the developers.

## *Rules for Collation of Responses*

Once the participatory exercise is completed in any one stage, the final step involves collating the responses that have been received and cohering them into a whole. Note that since the discussion prior has focused on stakeholder distribution, it is imperative that the collation of responses be carried out *stage wise*, and not at the end of the development of the AI program itself. In terms of the sieve, this means that the output from the 'sieve' is more like raw data, which must be acted upon by the developers and convert it into a more actionable form. That process of conversion is the subject of this sub-part.

Here too, as with stakeholder mapping and distribution, there is a significant gap in existing literature. As recently as in 2022, scholars have noted that 'gradually, however, we realised that the planners involved had a hard time explaining how the input was handled once gathered.'[105] They go further and describe the process of converting the input received from participants into actionable points as a 'black box'.[106] While some scholarship identifies that this is a necessary step in the overall participatory exercise, it does not go beyond into *how* collation ought to take place.[107]

Broadly, final collation can take place either through a voting mechanism or through consensus building generally.[108] Both mechanisms have certain advantages and disadvantages. In voting, provided that alternatives are delineated clearly, there is a conclusive answer as to what steps ought to be taken next. However, voting also suffers from various problems. First, a prior question that must be decided is whether everyone's votes ought to be counted equally, or whether votes should be weighted as per the importance of the stakeholder. In an AI which scans images for whether they show cancerous cells, the vote of medical professionals, by all counts, ought to count for more than, say, ethicists. Subjectivity in what weights ought to be given will, in most cases, be decided by the AI developers themselves. The second problem with voting is that while it may make sense with a group of trained experts (in the development stage, for instance), it may lead to undesirable outcomes where there is a lot of power imbalance among the participants. In such situations, it may make sense for the developers to gather all input first and use their judgement as to what should be incorporated.

---

[105] Erik Eriksson and others, 'Opening of the black box of participatory planning: a study of how planners handle citizens' input' (2022) 30(6) European Planning Studies 994, 995.
[106] ibid.
[107] Pieter-Jan Klok and Bas Denters, 'Structuring participatory governance through particular 'rules in use': lessons from the empirical application of Elinor Ostrom's IAD Framework' in Hubert Heinelt (eds.) Handbook on Participatory Governance (Elgar 2018).
[108] Fung (n 100); Klok and Denters (n 106).

The second method of collation is through a process of 'sorting', where the raw input is translated in a manner which is simplified into actionable points. This can take place in two ways- inclusive and selective sorting.[109] Both processes involve a great amount of professional judgement on the part of the planners.[110] In the former, no input is removed from consideration. Instead, a process of categorisation takes place as per urgency/importance, or as per any other criteria. In the latter, the developers will take calls on which input ought to be selected for consideration.[111]

| Voting | Advantages | <ul><li>Clear and actionable decisions, provided alternatives are delineated well.</li><li>Unless weighted voting is opted for, each person's preference gets counted equally.</li></ul> |
|---|---|---|
| | Disadvantages | <ul><li>Delineating alternatives relies upon subjective framing of the project managers.</li><li>Deciding the weightage of each vote increases subjectivity by the project managers. This will be the case even if the final decision is equal weightage to each vote.</li></ul> |
| Consensus Building | Advantages | <ul><li>Sorting allows for a more organic solution to be formed through deliberation among all parties.</li></ul> |
| | Disadvantages | <ul><li>Like voting, sorting also gives power to the project manager. An added disadvantage is that in selective sorting, some inputs may be opaquely removed from consideration entirely.</li><li>The process through which the input is sorted into an actionable decision will remain somewhat opaque.</li></ul> |

**Table 2:** *The advantages and disadvantages of voting and consensus building in the context of collating PAI responses.*

For input to be converted coherently, it will have to be translated from the language that the participants have used to language that can be used by the developers/planners. While there are different types of sorting, the core insight is that a level of translation is inevitable from what the participants say and what is encoded. For example, in the development of an AI software which screens mortgage applications and recommends their acceptance/denial, black applicants may give extremely detailed explanations as to why their applications are unfairly denied.[112] However, if this exercise is carried out at scale, then some level of simplification is necessary if the input is to also be made actionable.

---

[109] Eriksson and others (n 104).
[110] ibid 1000. "More voices are heard, and a variety of knowledges are considered, but when the participatory activities have been completed, the planners are usually left with voluminous input that they must handle based on professional judgment".
[111] ibid.

Developers will have to be mindful about the way in which this simplification and sorting takes place. Care must be taken to ensure that the material effect of the participatory input is not changed. This means that developers/project planners retain the power to decide which input is *incorporated*. This is precisely the disadvantage of the sorting process- while it allows for a more free-flowing discussion, and there is no aggregation of votes, it empowers developers/planners at the cost of participants. This increases the risk of tokenism, whitewashing, and the other dangers highlighted in prior sections. At the same time, the weightage of votes, and the framing of distinct issues, is also within the realm of discretion for developers.

Overall, the distinction between voting and sorting, and the advantages of one *vis-a-vis* and the other exists more within the realm of theory. The appropriate choice will depend upon the situation- the level of competence of participants, the time available, the type of input that is expected (if more detailed input is expected than sorting is a better option), etc.

---

[112] Kori Hale, 'A.I. bias causes 80% of black mortgage applicants to be denied' (Forbes 2 September 2021) <https://www.forbes.com/sites/korihale/2021/09/02/ai-bias-caused-80-of-black-mortgage-applicants-to-be-denied/?sh=58547e6f36fe>.

# Concluding Remarks

This paper has identified the primary reasons why a participatory approach in AI development can improve outcomes of the AI algorithm as well as enhance the fairness of the process. Even assuming this is so, success of the participatory exercise depends crucially on three factors- *firstly*, how the stakeholders are identified, *secondly*, how they are mapped across various stages of the AI lifecycle and *thirdly*, how their inputs are collated at each stage.

These three points are aptly summarised in the decision sieve. It is important to note that the diagram kicks into operation at *each point* where a decision must be made with participatory input. In other words, the process outlined in the diagram is an iterative process, performed each time such input is required. First, stakeholders are identified using the criteria highlighted above. Second, the 'sieve' of stakeholders for each decision is decided keeping in mind the factors of urgency, legitimacy, and power (in that order). Once this has been done, either before or after the decision is made on a particular point, the elicited input has to be collated either by voting (aggregation) or sorting mechanisms, as highlighted above. This translated input is forwarded to the next stage.

It is also worth noting that a decision point may both *require* horizontal or vertical translation and *necessitate* the same. The former will be the case where a decision point is only within a specific stage of the AI cycle, and the latter when it is from one stage to the next. Since the stages of the AI life cycle outlined above itself are not water-tight, a decision point may require both as well. This means that any decision taken, information given to participants as well as input elicited needs to be simplified and made communicable.

The next paper will deal with how these principles will be operationalised in two sectors- large language models in the healthcare sector and FRT in policing and law enforcement.

For any queries relating to this paper, please reach out to aditya.phalnikar@vidhilegalpolicy.in

www.vidhilegalpolicy.in

www.cerai.iitm.ac.in/

Vidhi Centre for Legal Policy
A-232, Defence Colony
New Delhi – 110024

Robert Bosch Centre for Data Science and AI, 5th floor, Block II, Bhupat and Jyoti Mehta School of Biosciences, Indian Institute of Technology Madras,
Chennai-600036

011-43102767 / 43831699

+914422574370